\documentclass[12pt,amsfonts]{article}
\usepackage{amsfonts}
\def\one{1\hskip-.37em 1}

\def\one{\mathbb{1}}

\def\half{{\textstyle{\frac{1}{2}}}}

\def\quarter{\textstyle{\frac{1}{4}}}

\def\p{\phi}

\def\hg{{\hat g}}
\def\hp{{\hat\pi}}
\def\de{\delta}

\def\l{\lambda}

\def\ra{\rightarrow}

\def\tint{{\textstyle\int}}
\def\hg{{\hat g}}
\def\hp{{\hat\pi}}

\def\s{\hskip.08em}

\def\d{\partial}

\def\b{\begin{eqnarray}}  
\def\e{\end{eqnarray}}    
\def\bn{\begin{eqnarray}}  
\def\en{\end{eqnarray}}   
\def\<{\langle}
\def\>{\rangle}

\def\no{\nonumber}

\def\de{\delta}

\def\{{\lbrace}
\def\}{\rbrace}

\def\one{1\hskip-.37em 1}

\def\half{{\textstyle{\frac{1}{2}}}}

\def\quarter{\textstyle{\frac{1}{4}}}

\def\p{\phi}

\def\hg{{\hat g}}
\def\hp{{\hat\pi}}

\def\l{\lambda}

\def\ra{\rightarrow}

\def\tint{{\textstyle\int}}
\def\hg{{\hat g}}
\def\hp{{\hat\pi}}

\def\s{\hskip.08em}
\def\d{\partial}

\def\b{\begin{eqnarray}}  
\def\e{\end{eqnarray}}    
\def\bn{\begin{eqnarray}}  
\def\en{\end{eqnarray}}   
\def\<{\langle}
\def\>{\rangle}
\def\d3{d^3\!x}

\def\no{\nonumber}

\def\de{\delta}

\def\{{\lbrace}
\def\}{\rbrace}

\def\half{\textstyle{\frac{1}{2}}}
\def\quarter{\textstyle{\frac{1}{4}}}

\def\p{\phi}

\def\tp{\tilde{p}}
\def\tq{\tilde{q}}
\def\tb{\tilde{\beta}}

\def\l{\lambda}

\def\t{\textstyle}

\def\ra{\rightarrow}
\def\tint{{\textstyle\int}}
\def\hg{{\hat g}}
\def\hp{{\hat\pi}}

\def\s{\hskip.08em}
\def\d{\partial}

\def\b{\begin{eqnarray*}}  
\def\e{\end{eqnarray*}}    
\def\bn{\begin{eqnarray}}  
\def\en{\end{eqnarray}}   

\def\<{\langle}
\def\>{\rangle}

\def\no{\nonumber}

\def\{{\lbrace}
\def\}{\rbrace}
\title{Enhanced Quantization:\\The {\it Right} way to Quantize {\it Everything}}
\author{John R. Klauder\footnote{Email: john.klauder@gmail.com}\\
Department of Physics and Department of Mathematics\\
University of Florida,
Gainesville, FL 32611-8440}
\begin{document}
\maketitle
\begin{abstract}Canonical quantization relies on Cartesian, canonical, phase-space coordinates to promote to Hermitian operators, which also become
the principal ingredients in the quantum Hamiltonian. While generally appropriate, this procedure can also fail, e.g., for covariant, quartic,
scalar fields in five-and-more spacetime dimensions (and possibly four spacetime dimensions as well), which become trivial; such failures
are normally blamed on the `problem' rather than on
the `quantization procedure'. In Enhanced Quantization
the association of $c$-numbers to $q$-numbers is chosen very differently such that: (i) there is no need to seek classical, Cartesian,
phase-space coordinates; (ii) {\it every} classical, contact transformation is applicable and {\it no} change of the quantum operators arises; (iii)
a new understanding of the importance of `Cartesian coordinates' is established; and
(iv) although discussed elsewhere in detail, the
procedures of enhanced quantization offer fully acceptable solutions yielding non-trivial results for quartic scalar fields in
four-and-more spacetime dimensions.   In early sections, this paper offers a  wide-audience approach to
the basic principles of Enhanced Quantization using simple examples;
later, several significant examples are cited for a deeper understanding. An historical note concludes the paper.
\end{abstract}

\section{Introduction}
Confirmation, by the outcome of an untold number of experiments, ensures the validity of the quantum theory as presently formulated.
Yet there are some troublesome cases, such as quantization of covariant, quartic, scalar fields in five and more spacetime dimensions,
or other nonrenormalizable examples, which do not
lead to acceptable results. Is it the `fault' of the problem itself, or is there something else going on? The purpose of this article is
to demonstrate that
a natural, but profound, change in how  a quantum theory and a classical theory are properly paired can lead to acceptable answers for those
problems mentioned above, and it can help many other problems as well.

The nature of the problem and the alternative procedures that make these good results occur is briefly outlined in the next two subsections.
Primary references for Enhanced Quantization are \cite{E1,E2,E3}.

\subsection{The problem}
Conventional quantization procedures take classical phase-space coordinates,
$p$ and $q$, with a Poisson bracket $\{q,p\}=1$,  and `promotes' them to Hermitian quantum operators, $P$ and $Q$, which obey
$[Q,P]\equiv QP-PQ=i\hbar\one$ as Hilbert-space operators.
There are many such `promotions'  that
are possible because of contact (a.k.a.~canonical) transformations, such as $\tilde{p}=\tilde{p}(p,q)$ and $\tilde{q}=\tilde{q}(p,q)$, which also obey
$\{\tilde{q},\tilde{p}\}=1$, and under `promotion', they also become Hermitian quantum operators, $\tilde{P}$ and $\tilde{Q}$, which
obey $[\tilde{Q},\tilde{P}]=i\hbar\one$ as well. Which pair of operators should be used in forming the quantum Hamiltonian operator? The
standard answer is
that the classical phase-space coordinates should be `Cartesian coordinates' \cite{dirac} (footnote, page 114). However, phase space does not have a metric to
decide which coordinates are Cartesian and which are not.
The usual kinetic energy term in a classical Hamiltonian, say, in a three-dimensional space,  can, up to a factor, be brought to the
form $p_1^2+p_2^2+p_3^2$,  which appears `Cartesian', so then such coordinates and their canonical partners (with no test of them
being `Cartesian') are chosen and, luckily, this generally  works very well. Moreover, if $\hbar>0$, but theoretically chosen to be extremely
tiny, then $P$ and $Q$ are still operators, i.e., $q$-numbers. For these terms to become $c$-numbers and commute, it is necessary that $\hbar=0$.
However, the classical world, as we encounter it,
requires that $\hbar>0$ and possibly even requires $\hbar$ to have its present value. Clearly, there is some uncertainty in choosing which are
the correct classical, canonical, phase-space variables to `promote' to quantum operators.

\subsection{The solution}
In the previous subsection we focused on {\it classical-to-quantum} connections; in this subsection we focus on {\it quantum-to-classical} connections.
Important features of this subsection are the fact that $\hbar$ is allowed (i) to retain its true positive value and (ii) to highlight its
role as an important parameter.

The quantum action functional is given by
  \bn A_Q=\tint_0^T\,\<\psi(t)|\s[i\hbar(\d/\d t)-{\cal H}(P,Q)\s]\s|\psi(t)\>\,dt \label{e123}\;, \en
and general stationary variations of normalized vectors $\{|\psi(t)\>\}$, holding $|\psi(T)\>$ and $|\psi(0)\>$ fixed, lead to Schr\"odinger's equation
   \bn i\hbar\,\d\s|\psi(t)\>/\d t={\cal H}(P,Q)\,|\psi(t)\> \en
   as  well as its adjoint. But suppose that your variations were limited to a set of two-parameter vectors of the form
   \bn  |p,q\>\equiv \exp[-iqP/\hbar]\,\exp[ip\s Q/\hbar]\,|0\>\;, \label{e335} \en
   where, for convenience, we have chosen the `fiducial vector' $|0\>$ as a normalized solution of the equation $(Q+iP)|0\>=0$; note that we
   have omitted any other dimensional factor and thus both $Q$ and $P$ (as well as $q$ and $p$), effectively, have dimensions of $\hbar^{1/2}$.
   Moreover, both $Q$ and $P$, which obey $[Q,P]=i\hbar\one$, are chosen self adjoint, a criterion much stronger than Hermitian, which ensures
   that the two exponential terms
    in (\ref{e335}) are unitary, and therefore every vector $|p,q\>$, with
   $(p,q)\in \mathbb{R}^2$, is normalized.  This set of vectors is also one example of a set of canonical `coherent states' \cite{BSJK}, and we shall
   sometimes refer to them by that name.

   Now, let us assume that the domain of the quantum action functional only contains the set of vectors $\{|p,q\>\}$, which then leads to a restricted (R),
   quantum action functional given, with $\dot{q}(t)=d q(t)/d t$, by
     \bn && A_{Q(R)}=\tint_0^T\,\<p(t),q(t)|\s[i\hbar(\d/\d t)-{\cal H}(P,Q)\s]\s|p(t),q(t)\>\,dt\no\\
          &&\hskip2.85em =\tint_0^T\,[\s p(t)\s \dot{q}(t)- H(p(t),q(t))\s]\, dt \;. \label{e233}\en
   Clearly, this result looks exactly like a classical action functional, and it has several very interesting features:  (i) the value of $\hbar$ is
    never changed from its natural value, and $\hbar$ may appear in the equations of motion. For this reason we call Eq.~(\ref{e233}) the
    {\it enhanced classical action functional}; (ii) both $p$ and $q$ are simultaneously $c$-numbers; (iii) the expression
    $H(p,q)\equiv \<p,q|\s{\cal H}(P,Q)\s|p,q\>$ is called the Weak Correspondence Principle \cite{WCP}, and it follows, e.g., for a
    polynomial ${\cal H}(P,Q)$,
      that $H(p,q)=\<0|\s{\cal H}(P+p\one,Q+q\one)\s|0\>={\cal H}(p,q)+{\cal{O}}(\hbar;p,q)$. Observe that the quantum function ${\cal H}$
      equals the classical function $H$, up to terms in $\hbar$, which is {\it exactly} what is sought for by seeking `Cartesian coordinates'
      in canonical quantization procedures.
    Although phase space has no metric, Hilbert space has one and that metric, relevant for ray vectors, leads to (a multiple of) the
    Fubini-Study metric \cite{FS} for two, infinitely
     close, coherent-state ray vectors given by
       \bn d\sigma^2\equiv 2\hbar\s[\,\|\,d\s|p,q\>\s\|^2-|\s\<p,q|\,d|p,q\>\s|^2\s]=dp^2+dq^2 \;,\en
      an expression which gives a whole new meaning to `Cartesian coordinates'\footnote{The simple form of $d\sigma^2$ owes much to the
      equation $(Q+iP)\s|0\>=0$ that defines the fiducial vector, but using a general, normalized, fiducial vector $|\eta\>$, it follows
      that $d\sigma^2=(2/\hbar)[A\s dp^2+B\s dp\s dq+C\s dq^2]$,
      where $A=\<(\Delta Q)^2\>$, $B=\<\s(\Delta Q\s\Delta P+\Delta P\s\Delta Q)\s\>$, and $C=\<(\Delta P)^2\>$, with $\<(\cdot)\>\equiv\<\eta|(\cdot)
      |\eta\>$ and $\Delta X\equiv X-\<X\>$. Clearly, a suitable linear change of the coordinates would lead to Cartesian coordinates.}; and
    (iv) a point in phase space may be described by $p$ and $q$ in one coordinate system, and the same point may be
    described by $\tp$ and $\tq$ in another coordinate system. Contact transformations limit such coordinate transformations so that
    these variables also satisfy the one form $p\s\s dq=\tp\s\s d\tq+d{\tilde G}(\tp,\tq)$.
    The map of phase-space points into Hilbert-space vectors therefore requires that $|\tp,\tq\>\equiv|p(\tp,\tq)),q(\tp,\tq)\>=|p,q\>$ since the change of
    coordinates still must map the same phase-space point into the same vector in Hilbert space. Consequently, it follows that
     \bn && A_{Q(R)}=\tint_0^T\,\<\tp(t),\tq(t)|\s[i\hbar(\d/\d t)-{\cal H}(P,Q)\s]\s|\tp(t),\tq(t)\>\,dt\no\\
          &&\hskip2.85em =\tint_0^T\,[\s \tp(t)\s \dot{{\tq}}(t)+\dot{{\tilde G}}(\tp(t),\tq(t))- \tilde{H}(\tp(t),\tq(t))\s]\, dt \; \en
   which leads to a proper change of classical canonical coordinates {\it without disturbing the quantum operators whatsoever}.

 \subsection{Discussion}
    It is our belief that the several significant benefits that follow from Eqs.~(\ref{e123}) and (\ref{e233}) as outlined above represent the correct
    interpretation of the classical version of a quantum system\footnote{The term ${\cal{O}}(\hbar;p,q)$ could (i) significantly modify the
    nature of classical behavior, or (ii) in other systems in which different fiducial vectors occur, it could account for some ambiguity in the
    enhanced classical dynamical behavior; but since such terms arise at the quantum level, they are negligible for any macroscopic system}.
    Moreover, by finding phase-space coordinates that serve as `Cartesian coordinates', which
    are identical with such coordinates from a canonical quantization viewpoint, it follows that {\it Enhanced Quantization results agree
     with Canonical Quantization results, when the latter lead to acceptable results}.

\subsection{Some physics}
The previous subsection offered some interesting mathematics about what coherent states $\{\s|p,q\>\s\}$ could offer for a different
connection between $c$-numbers and $q$-numbers, but it offered no convincing physical argument that this subset of vectors was the `right choice'.
In the present subsection we argue that this set of coherent states is ideally suited to the task.

A macroscopic (i.e., classical) observer of a microscopic system must be sure that any observations s/he makes must not disturb the system.
On the face of it, this sounds like an impossible task. However, there are a few ways in which this can be done. As a real world example
of a reliable---but indirect---measurement of the height of a very tall pole, recall the procedure of measuring the length of the shadow of
the pole and the shadow length and real length of a much smaller object, and you have the data to determine the height of the tall pole.
For our microscopic system, we start with the knowledge that it is a quantum system and thus has many quantum states. We postulate that the
wave function $\eta(x)$, where $x$ is a coordinate variable,
 is one of the wave functions in the relevant $L^2( \mathbb{R})$ Hilbert space. To measure $\eta(x)$ all we need to do is find how that function
 varies in space! A displacement of the system along the $x$ axis by an amount $q$ leads to the function $\eta(x-q)$. However, we need not
 move the microscopic system but rather move the observer a comparable distance in the opposite direction, and thanks to Galilean invariance,
 the result is the same:  $\eta(x-q)$. In so doing we can map out the function theoretically, and with that knowledge, we can gain alternative
 realizations of that function such as its Fourier transformation, among other possibilities. Indeed, the Fourier transform is also available
 to us if the system moves at a constant velocity of varying amount.\footnote{To appreciate the effect of motion in shifting the Fourier
 transform, one may recall that Christian Doppler hired a train, put a musical band on a flat car, and ran the train through a terminal
 where spectators clearly heard the band have a higher pitch as they approached and a lower pitch as they departed.} But once again it is
 not necessary to move the microscopic system at a constant velocity because Galilean invariance also states that we get the same effect if
 we put the observer in motion at the same speed in the opposite direction. Thus we have a two-parameter family of states that are part
 of the system's quantum states and we have arrived at that information without ever touching the microscopic system. If we put this information
 into quantum mechanical language in the $x$ representation we get $e^{ip(x-q)/\hbar}\s\eta(x-q)$, which is just the $x$ representation of the
 abstract vector $\exp[-iqP/\hbar]\s \exp[ipQ/\hbar]\s|\eta\>$.
Summarizing, the meaning of the parameter $q$ is, as is clear from the $x$ representation,  a coordinate variable, while the parameter $p$ is a momentum
 variable related to a velocity variable based on the relation $\dot{q}=\d H(p,q)/\d p=F(p)$, which holds for many systems.

 But, hold on, not all systems have the property that $\dot{q}=\d H(p,q)/\d p=F(p)$. What happens then?

\section{Affine Variables}

Consider the classical action functional given by
  \bn A_C=\tint_0^T\s[p(t)\dot{q}(t)-q(t)p(t)^2 ]\,dt\;, \en
  with the requirement that $q(t)>0$.\footnote{This example is a toy model of gravity where $q>0$ plays the role of the metric tensor with its
  positivity constraint while $p$ plays the role of minus the Christoffel symbol \cite{AK2}; note: this reference uses different notation.} In this case, $\dot{q}(t)=2q(t)p(t)$ and thus a constant velocity does
   not mean a constant momentum, but, more importantly, $q(t)>0$ means that $q$ {\it cannot} be a `Cartesian variable'.

   This model is simple enough to solve, and we find that $q(t)=q_0\s (1+p_0\s t)^2$ and $p(t)=p_0/(1+p_0\s t)$, and thus if the energy $q_0\s p_0^2>0$
   there is a singularity, $q(-1/p_0)=0$, in the solution. Does quantization remove the singularity?
\subsubsection{Are canonical variables available?}
   One common reaction to this
   problem is to change variables so
   that  $q(t)=\exp[2\s x(t)]$ and thus while $q\in \mathbb{R}^+$, we find that $x\in \mathbb{R}$. It follows that $\dot{q}(t)=2\s\dot{x}(t)\s q(t)$. Moreover,
   the new momentum $p_x(t)=2q(t) p(t)$,
    and thus the new version of the classical action functional becomes
      \bn A_C=\tint^T_0\s[ p_x(t) \dot{x}(t)-\quarter p_x(t)^2\s e^{-2x(t)}\s]\,dt \;.\en
   Observe that a singularity occurs when $q=0$ and $p=\infty$, or when $x=-\infty$ and $p_x=0$.
 While $\dot{x}\not=\tilde{F}(p_x)$, we have achieved canonically conjugate variables such that $\{x,p_x\}=1$, and they {\it might} be
 `Cartesian variables'. With fingers crossed, these variables
 are promoted to quantum operators $X$ and $P_x$ for which $[X,P_x]=i\hbar\one$, and they enter the Hamiltonian operator ${\cal H}=P_x\s e^{-2X}\s P_x$
 as usual.  This procedure generates a quantum story, but is it the `correct story'?

  \subsection{A new pair of operators}
  Consider a canonical pair of irreducible operators $Q$ and $P$ that satisfy $[Q,P]=i\hbar\one$. Multiplication by $Q$ leads to $  Q\s[Q,P]
  =[Q,QP]=[Q,\half(QP+PQ)]= [Q,D]=i\hbar Q$, where $D\equiv\half(QP+PQ)$. These equations describe the Lie algebra for the {\it affine variables},
  $D$ and $Q$.
  While $P$ acts to {\it translate} $Q$, $D$ acts to {\it dilate} $Q$. This leads to two principal irreducible representations, one with $Q>0$ and
  the other by $Q<0$ \cite{RU2,AK3}; a third representation has $Q=0$, but it is
  less important. For the choice $Q>0$ (chosen dimensionless as is $q$), these operators define the affine coherent states \cite{BSJK}
    \bn |p,q\>=\exp[ip\s Q/\hbar]\s\exp[-i\ln(q)D/\hbar]\s|\tb\>\;,  \en
   where $(p,q)\in  \mathbb{R}\texttt{x} \mathbb{R}^+$, a domain that helps distinguish the affine coherent states in this section from the canonical coherent states.
   The fiducial vector $|\tb\>$ is chosen as a normalized solution of $[(Q-1)+iD/\tb\s]\s|\tb\>=0$, with an $x$ representation given by
   $\tb(x)= M\s x^{\tb/\hbar-1/2}\s e^{-\tb\s x/\hbar}$, with  $\tb>0$, which also leads
   to $\<\tb|Q|\tb\>=1$, $\<\tb|Q^n|\tb\>=1+{\cal O}(\hbar)$ for $n\ge2$, and $\<\tb|D|\tb\>= 0$.
   Given the quantum action functional
     \bn A_Q=\tint_0^T \s\<\psi(t)|\s[i\hbar(\d/\d t)-{\cal{H}}'(D,Q)\s]\s|\psi(t)\>\,dt\;, \en
     we readily find the enhanced classical action functional given by
     \bn &&A_{Q(R)}=\tint_0^T \s\<p(t),q(t)|\s[i\hbar(\d/\d t)-{\cal{H}}'(D,Q)\s]\s|p(t),q(t)\>\,dt\no\\
            &&\hskip2.7em=\tint_0^T [-q(t)\s \dot{p}(t)-H(p(t),q(t))\s]\,dt\;. \en
     Clearly, this equation describes a classical canonical system. Moreover, the weak correspondence principle shows that
      \bn &&H(p,q)\equiv H'(pq,q)=\<p,q|\s{\cal H}'(D,Q)\s|p,q\> \no \\
             &&\hskip8.6em  = \<\tb|\s{\cal H}'(D+pqQ,qQ)\s|\tb\> \no\\
            &&\hskip8.6em  =  {\cal H}'(pq,q)+{\cal O}(\hbar;p,q) \;. \en
  The true classical Hamiltonian $H_c(p,q)=\lim_{\hbar\ra0}\s {H}'(pq,q) ={\cal H}'(pq,q)$.

  For the toy model that introduced this section, we can recast the quantum Hamiltonian into affine quantum variables as ${\cal H}'(D,Q)=DQ^{-1}D$. In that
  case, the enhanced classical Hamiltonian becomes
    \bn    && H(p,q)=\<\tb| (D+pqQ)(qQ)^{-1}(D+pqQ)\s|\tb\>\no \\
            &&\hskip3.368em = qp^2+\<\tb|D(qQ)^{-1}D|\tb\>\no \\
             &&\hskip3.358em = qp^2+ \hbar^2C'/q\;, \en
    where $C'>0$. This result has a profound influence on the classical dynamics and the enhanced classical dynamics: If $\hbar=0$ the classical
    solutions with positive energy all encounter a
    singularity where $q=0$. However, thanks to the term $\hbar^2 C'/q$, solutions of the enhanced classical dynamics
    do {\it not} encounter singularities.

    The variable $q>0$ acts to dilate the coordinate $x$ as in the $x$ representation of the affine coherent states given by
    $q^{-1}\s e^{ipx/\hbar}\s \tb(x/q)$.
    The variable $p$ still refers to the Fourier transformation, while $\tb(x/q)$ is another construction
    arising from the knowledge of the translated function $\tb(x)$. The variables $(p,q)$ can not be Cartesian, so what are they?
    The (scaled) Fubini-Study metric for the affine coherent states provides an answer:
        \bn d\sigma^2\equiv 2\hbar\s[\,\|\,d\s|p,q\>\s\|^2-|\s\<p,q|\,d|p,q\>\s|^2\s]= \tb^{-1} q^2\s dp^2+\tb\s q^{-2}\s dq^2\;, \en
    which is the metric of a space of constant negative curvature: $-2/\tb$ that is geodetically complete. In this sense, the affine coherent states
    provide a vast number of reference frames for different $\tb$ values, along with the canonical coherent states
    which form a flat space.

\section{Spin Variables}
For completeness regarding conventional two-parameter coherent states, we include a brief discussion of spin coherent states. We are given three, irreducible  spin
operators which obey $[S_1,S_2[=i\hbar\s S_3$ and cyclic permutations, where $\Sigma_{j=1}^3\,S_j^2=\hbar^2\, s(s+1)\one_s$, with a Hilbert space
dimension of $(2s+1)$ and spin value $s\in\{1/2,1,3/2,\cdots\}$. The normalized eigenvectors of $S_3$
satisfy $S_3\s |s,m\>=m\s\hbar\s|s,m\>$, with $-s\le m \le s$; the fiducial vector is chosen as $|s,s\>$, which is a normalized solution of the equation
$(S_1+iS_2)\s|s,s\>=0$. Finally, the spin coherent states are given \cite{BSJK} by
   \bn |\theta, \phi\>\equiv e^{-i\phi\s S_3/\hbar}\,e^{-i\theta\s S_2/\hbar}\,|s,s\>\;, \en
   where $0\le \theta\le\pi$ and $0\le\phi<2\pi$.

   We next introduce the quantum action functional
      \bn A_C =\tint_0^T \<\psi(t)\s[\s i\hbar(\d/\d t)-{\cal H}({\bf S})\s]\s|\psi(t)\>\,dt \;, \en
  and a restricted quantum action functional using spin coherent states  leads to
    \bn &&A_{Q(R)}=\tint_0^T \<\theta(t),\phi(t)|\s[\s i\hbar(\d/\d t)-{\cal H}({\bf S})\s]\s|\theta(t),\phi(t)\>\, dt\no\\
         &&\hskip2.8em = \tint_0^T\s[\s s\hbar\cos(\theta(t))\,\dot{\phi}(t)-H(\theta(t),\phi(t))\s]\, dt\;.\en
         To look more like the discussion in earlier sections, we can introduce $p\equiv (s\s\hbar)^{1/2}\s \cos(\theta)$ and
         $q\equiv (s\s\hbar)^{1/2}\s \phi$. Thus we have $ |p,q\> \equiv |\theta,\phi\>$, and the enhanced classical action functional
         becomes
          \bn A_{Q(R)}=\tint_0^T\s [\s p(t)\dot{q}(t)-H(p(t),q(t))\s]\,dt \;,\en
          where $- (s\s\hbar)^{1/2}\le p\le (s\s\hbar)^{1/2}$ and $-\pi (s\s\hbar)^{1/2}< q\le \pi (s\s\hbar)^{1/2}$.

          Not surprisingly, the  (scaled) Fubini-Study metric leads to
          \bn d\sigma^2\equiv 2\hbar\s[\,\|\,d\s|p,q\>\s\|^2-|\s\<p,q|\,d|p,q\>\s|^2\s]=(s\hbar)[\s d\theta^2+\sin(\theta)^2\,d\phi^2\s] \;,\en
          which describes a spherical surface with a constant positive curvature: $(s\hbar)^{-1}$, and only for those $s$ values that are allowed.

           Of course, nothing remains in this section if $\hbar\ra0$.

\section{The Power of Enhanced Quantization}
In the following subsections we outline several problems that lead to unsatisfactory results when treated by canonical quantization and, instead, lead to
satisfactory results when treated by enhanced quantization. References to full treatment of these examples are offered and the reader is urged to
tackle these problems by standard methods to appreciate the difference in the two approaches. A reference that treats all of these examples is \cite{E3},
and to a lesser extent \cite{E2}.

\subsection{Rotationally symmetric models}
This example has a classical Hamiltonian given, for $0<m_0<\infty$ and $0\le g_0<\infty$,  by the expression
  \bn H(p,q)=\Sigma_{n=1}^N \s[\s p^2_n+m_0^2\s q^2_n\s]+ g_0\s \{\Sigma_{n=1}^N q^2_n\s\}^2\;, \en
  where $N\le\infty$,   $p=\{p_1,p_2,\cdots\}$, and $q=\{q_1, q_2,\cdots\}$; if $N=\infty$ it is necessary that allowed sequences obey
  $\Sigma_{n=1}^\infty\s[\s p^2_n+q^2_n\s]<\infty$. The Poisson bracket for these variables is given by $\{q_m, p_n\}=\delta_{mn}$. Other
  powers of the interaction term follow a similar analysis.

  The classical solutions to this problem have a `shuffle  symmetry'. To illustrate this symmetry, let case A have initial conditions
  that are all zero except
  for, say, three different $n$ numbers. For example, in case A let the
  three variables for $n=1, 2, 3$ have nonzero initial conditions, while in case B the variables $n=4, 5, 6$ have the same initial conditions,
  or, in fact, any other three variables such as $n=7, 19, 125$, etc. The solution for each of these cases is identical and illustrates
  shuffle  symmetry.
  Canonical quantization, for $N=\infty$, leads to a trivial (= free) quantum theory result.  Such a result at least obeys quantum shuffle  symmetry, but of course, it
  is otherwise unfaithful since we started with a non-free ($g_0>0 $) classical model and, after letting $\hbar\ra0$, end up
  with a free ($g_0=0$) classical model.

  Enhanced quantization yields a non-free quantum model that also enjoys shuffle  symmetry. This solution uses {\it reducible} representations of
  the basic operators which is permitted by the weak correspondence principle $H(p,q)=\<p,q|\s{\cal H}(P,Q,\cdots)\s|p,q\>$.

  Principal references
    for rotationally symmetric models are \cite{RS1,BCQ,E2, E3}.

\subsection{Ultralocal scalar fields}
The classical action functional for an ultralocal quartic scalar model is given (for $s$ spatial dimensions) by
  \bn A_C=\tint_0^T\tint \{\,\half[\dot{\phi}(x,t)^2-m_0^2\s\phi(x,t)^2\s] -\l_0\s\phi(x,t)^4\s\}\,d^s\!x\,dt\;, \en
  which contains the time derivative of the field but no spatial derivatives. Although we focus on a quartic interaction, the treatment of
  other nonlinear terms follow a similar story, so long as they are lower bounded.

  A canonical quantum treatment often begins by first replacing the spatial continuum by a finite spatial lattice thus reducing the problem to a
  discrete, finite number of identical, independent, non-trivial, quantum-mechanical models. These models can be separately solved, and the next
  step  to complete the quantization is to take the continuum limit of the spatial lattice. Ultimately, the Central Limit Theorem controls the
  continuum limit, and these models merge into a Gaussian ground-state distribution signaling a trivial (= free) solution. Again, we find that a non-free
  classical model has become a free quantum model, which, as $\hbar\ra0$, implies a free classical behavior in contrast with the starting model.

  This nonrenormalizable model can be solved by enhanced quantization techniques that yield an acceptable, nontrivial quantum theory.  
  The solution depends on an unusual, $\hbar$ dependent, counter term, created from the  basic ingredients that include
  reducible affine quantum field operators, and leads to the other behavior of the Central Limit Theorem: a generalized Poisson
  distribution for the ground-state distribution. This form of the quantum solution leads back to the initial, non-linear,
  classical model, when $\hbar\ra0$.

  Principal references for ultralocal scalar models are \cite{UL,UL2,BCQ,E2,E3}.

\subsection{Covariant scalar field}
The classical action functional for a quartic, covariant scalar field is given (for $x\in \mathbb{R}^s$) by
  \bn A_C=\tint_0^T\tint\,\{\,\half[\dot{\phi}(x,t)^2-(\overrightarrow{\nabla}{\phi})(x,t)^2-m_0^2\s\phi(x,t)^2\s] -\l_0\s\phi(x,t)^4\s\}\,d^s\!x\,dt\,. \en
With the spacetime dimension $n \equiv (s+1)$, these models are denoted by $\phi^4_n$; a similar analysis for $\phi^p_n$, $p\in\{6, 8, 10,\cdots\}$,
may be considered as well. In Euclidean coordinates, where $\phi(x,t)\ra\phi(x)$ with
the second $x\in \mathbb{R}^n$, there is an important multiplicative inequality \cite{RU2, BCQ}
regarding such models which reads
\bn  \{\tint \phi(x)^4\,d^n\!x\}^{1/2}\le C_{n}\,\tint [\s(\nabla{\phi})(x)^2+m^2_0\,\phi(x)^2\s]\,d^n\!x\;, \label{E333}\en
where $C_{n}=(4/3)\s m_0^{(n-4)/2}<\infty$ for $n\le4$, and $C_{n}=\infty$ for $n\ge5$. The latter case implies that there are
fields, e.g., $\phi_{sing}(x)=|x|^{-\alpha}\s e^{-x^2}$, where $n/4\le \alpha<(n-2)/2$,  for which the integral on the left diverges while
the integral on the right is finite. It is important to note that for $n\ge5$ the domain of the classical action does not equal the domain
of the free model, and thus,
as $\l_0\ra0$, the limiting domain is smaller than the domain of the free model. These are cases in which the interacting
models are {\it not} continuously connected to their own free theory, and which we refer to as `pseudofree models'.

Canonical quantization of these models provides acceptable results for $n=2,3$ \cite{GJ}. For $n=4$, renormalization group
calculations \cite{DC} as well as Monte Carlo
simulations \cite{FW} point toward triviality, while for $n\ge5$ triviality has been proved \cite{MA,JF}.  In fact, the inequality (\ref{E333}) also
distinguishes the {\it renormalizable} quantum models when $n\le4$ and the {\it nonrenormalizable} quantum models when $n\ge5$ \cite{BCQ,E3}.
 These results reflect the present understanding based on standard canonical quantization procedures.

The fact that for $n\ge5$ the interacting classical model is not connected to its own free model strongly suggests that for $n\ge5$ the interacting quantum model
is not connected to its own free quantum model. A similar situation arose for the ultralocal scalar models discussed above, and this led to
a pseudofree quantum ground state. Fortunately, a natural modification of the ultralocal model pseudofree ground state serves to be the
right choice for the covariant scalar pseudofree model. In a similar manner the basic ingredients for $n\ge5$ include reducible affine field operators,
and lead to an unusual counter term that resolves all problems for the nonrenormalizable case when $n\ge5$. Although the $n=4$ model is not
a nonrenormalizable model, we can nevertheless extend the unusual counter term to lower spacetime dimension models, which leads to
alternative models for $n=2,3,4$ \cite{PP}. The analysis for this potentially new $n=4$ case involves Monte Carlo simulations. In fact, a preliminary study
\cite{JS} suggests that
for $n=4$ the new model is {\it non-free} as determined by a nonvanishing renormalized coupling constant.  For those cases where
$n=2, 3$ these results confirm the
fact that more than one kind of renormalization is possible in these cases.

Primary references for quartic, covariant, scalar fields are
\cite{JK1,JK2,PP,E2,E3}.

\subsection{Affine quantum gravity}
 The ADM \cite{ADM} formulation of the classical action functional involves
phase-space variables, namely the spatial momentum $\pi^{cd}(x,t)\;[=\pi^{dc}(x,t)\s]$ and the spatial metric $g_{ab}(x,t)\;[=g_{ba}(x,t)\s]$,
 with $a,b,c,d\in\{1,2,3\}$, the latter of which must, for all $(x,t)$, fulfill the {\it metric-positivity requirement}: $u^a\s g_{ab}(x,t)\s u^b>0$
 (summation convention here and below) for any $\Sigma_a\s (u^{a})^2>0$. The classical action functional is given by
\bn &&A_C=\tint_0^T\tint\{\s -g_{ab}(x,t)\s\dot{\pi}^{ab}(x,t)-N^a(x,t)\s\s H_a(\pi,g)(x,t)\no\\
    &&\hskip9em-N(x,t)\s\s H(\pi,g)(x,t)\s\}\,d^3\!x\,dt\;, \en
   where $N^a(x,t)$, and $N(x,t)$ are Lagrange multiplier fields, $H_a(\pi,g)(x,t)=-2\pi^b_{a\s|\s b}(x,t)$, with
   $\pi^b_a(x,t)\s\equiv \pi^{bc}(x,t)\s g_{ca}(x,t)$,
   are the classical
   diffeomorphism constraints, where $_|$ denotes a covariant derivative using the spatial metric alone, and  $H(\pi,g)(x,t)$ is
   the classical Hamiltonian constraint, where
 \bn &&H(\pi,g)(x,t)=g(x,t)^{-1/2} \,[\s\pi^a_b(x,t)\s\pi^b_a(x,t)-\half\s\pi^a_a(x,t)
  \s\pi^b_b(x,t)\s] \no\\
&&\hskip8em   +g(x,t)^{1/2}\, R(x,t) \;; \label{h45} \en
here $R(x,t)$ denotes the three-dimensional, spatial scalar curvature.

Canonical quantization of this system promotes $g_{ab}(x,t)$ to $\hg_{ab}(x,t)$ and preserves the  metric-positivity requirement both as
a $c$-number as well as a $q$-number, but this requirement forces $\hp^{cd}(x,t)$ to be Hermitian and {\it not} self adjoint, much as we saw
in the toy model of gravity in Sec.~2. Other approaches can realize the metric as the square of some other variable, pictorially speaking, but
that other variable can vanish which then breaks the  metric-positivity requirement.

In an enhanced quantization approach we are drawn to affine variables that can have locally self-adjoint operators as well as respect the
metric-positivity requirement. The basic affine variables involve
 $\pi^a_b(x)$ and $g_{ab}(x)$, and it is these  variables that are promoted to quantum operators in the affine
 quantum gravity program \cite{G1}; specifically  (with $\hbar=1$),
   \bn &&[\hp^a_b(x),\hp^c_d(y)]=i\half\s[\de^c_b\,\hp^a_d(x)-\de^a_d\,\hp^c_b(x)]\,\de(x,y)\;,\no\\
    && \hskip-.08cm[\hg_{ab}(x),\hp^c_d(y)]=i\half\s[\de^c_a\,\hg_{db}(x)+\de^c_b\,\hg_{ad}(x)]\,\de(x,y)\;,\\
  && \hskip-.18cm[\hg_{ab}(x),\hg_{cd}(y)]=0\;\no  \label{e1}\en
between the metric components $\hg_{ab}(x)$ and the components of the mixed-index momentum (also referred as the `momentric')
field operator $\hp^a_b(x)$, the
quantum version of the classical variable $\pi^a_b(x)\equiv \pi^{ac}(x)\s g_{cb}(x)$.

The affine coherent states are defined (for $\hbar=1$) by
  \bn |\pi,\gamma\>\equiv e^{\t i\tint \pi^{ab}(x)\hg_{ab}(x)\,d^3\!x}\,e^{\t-i\tint\gamma^a_b(x)\hp^b_a(x)\,d^3\!x}
  \,|\eta\>\,\;\;[\,= |\pi,g\>\,] \label{e2}\en
for general, smooth, $c$-number fields $\pi^{ab}(x)\,[=\pi^{ba}(x)]$ and $\gamma^c_d(x)$ of compact support, and the
fiducial vector $|\eta\>$ is chosen so that the coherent-state overlap functional becomes
 \bn  &&\hskip-.5cm\<\pi'',g''|\pi',g'\>\equiv \exp\bigg(\!-\!2\int b(x)\,d^3\!x\,  \label{e3} \\
  &&\hskip-.1cm\times\ln\bigg\{  \frac{
\det\{\half[g''^{kl}(x) +g'^{kl}(x)]+i\half b(x)^{-1}[\pi''^{kl}(x)-
\pi'^{kl}(x)]\}} {(\det[g''^{kl}(x)])^{1/2}\,(\det[g'^{kl}(x)])^{1/2}}
\bigg\}\bigg) \;.\no \en
Observe that the matrices $\gamma''$ and $\gamma'$ do {\it not} explicitly appear in (\ref{e3}) because the
 choice of $|\eta\>$ is such that each $\gamma=\{\gamma^a_b\}$ has  been replaced by $g=\{g_{ab}\}$, where
  \bn  g_{ab}(x)\equiv [e^{\t\gamma(x)/2}]_a^c\,\<\eta|\hg_{cd}(x)|\eta\>\,[e^{\t\gamma(x)/2}]_b^d \;.\en
Note that the functional expression in (\ref{e3}) is ultralocal, i.e., specifically of the form
  \bn  \exp\{-\tint b(x)\,d^3\!x\,L[\pi''(x),g''(x);\pi'(x),g'(x)]\,\}\;, \en
 and thus, there are no correlations between spatially separated field values, a neutral position adopted towards
 spatial correlations before any constraints  are introduced \cite{G2}.  On invariance  grounds, (\ref{e3}) necessarily involves a
 {\it scalar density} $b(x)$, $0< b(x)<\infty$, for all $x$; this arbitrary and non-dynamical auxiliary function $b(x)$,
 with dimensions (length)$^{-3}$,
   should disappear when the gravitational constraints are fully enforced, at which point proper field correlations will
   arise. In addition, note that the coherent-state overlap functional is {\it invariant} under general spatial
   coordinate transformations. Finally, we emphasize that the expression $\<\pi'',g''|\pi',g'\>$ is a {\it continuous functional
   of positive type} and thus may be used as a {\it reproducing kernel} to define a {\it reproducing kernel Hilbert space}
(see \cite{RK})  composed of continuous phase-space functionals $\psi(\pi,g)$ on which elements of
the initial, ultralocal representation of the affine field operators act in a natural fashion.

Thanks to the choice of the fiducial vector $|\eta\>$ defining the affine coherent states, the coherent states have a complex polarization, which leads
to the coherent-state overlap function
(\ref{e3}) having a functional-integral representation involving a well-defined probability measure \cite{TT} (Chapter 8).
Moreover, the coherent-state overlap function, and its representation via a functional integral, serve as a reproducing kernel for the kinematical
Hilbert space prior to introducing the four constraint fields that describe general relativity.
The introduction of the constraint fields into the coherent-state functional integral by the {\it projection operator method} \cite{TT} (Chapter 9),
which is also able to accommodate
second-class quantum constraints (which quantum gravity has),  implicitly leads to a
reproducing kernel that represents the physical Hilbert space for quantum gravity \cite{G5}. Although this latter integral representation appears too
complicated for an analytic evaluation, it may be approximately calculated numerically.


Primary references for affine quantum gravity are \cite{G1,G2,G3,G4,G5,E2,E3}.

\section{Historical Note}
As a natural development of the author's 1959 thesis (published in 1960) \cite{TH}, the author first found Eq.~(\ref{e233}) in 1962 \cite{BB} and assumed it was `coincidental'
 in relation to the `genuine' procedures of canonical quantization.
However, over the years, and still maintaining this coincidental view, the weak correspondence principle  was used to solve several difficult problems.
Only in 2012 did the author finally
accept that the principles of
enhanced quantization are the `correct way' to link a quantum model and a classical model, which opened the door to a general
application of these alternative quantization procedures. We encourage others to apply
 the procedures of enhanced quantization to their own problems.
\section*{Acknowledgements}
The author thanks T. Adorno, J. Ben Geloun, and  G. Watson for their contributions to the enhanced quantization program and its consequences.



\begin{thebibliography}{99}  
\bibitem{E1}J.R. Klauder, ``Enhanced Quantization: A Primer'', J. Phys. A: Math. Theor. {\bf 45},  285304 (8pp) (2012); arXiv:1204.2870.
\bibitem{E2} J.R. Klauder, ``Enhanced Quantum Procedures that Resolve Difficult Problems'', Rev. Math. Phys. 27 (2015) no.05, 1530002 (43pp):
 arXiv:1206.4017.
\bibitem{E3} J.R. Klauder, {\it Enhanced Quantization: Particles, Fields \& Gravity}, (World Scientific, Singapore, 2015).
\bibitem{dirac} P.A.M. Dirac, {\it The Prnciples of Quantum Mechanics}, (Clarendon Press, Oxford, 1958).
\bibitem{BSJK}  J.R. Klauder and B.-S. Skagerstam, {\it Coherent States:  Applications to Physics and Mathematical Physics}, editors plus an original
introduction (World Scientific, Singapore, 1985).
\bibitem{WCP}  J.R. Klauder, ``Weak Correspondence Principle", J. Math. Phys. {\bf 8}, 2392-2399 (1967).
\bibitem{FS} See, e.g., http://en.wikipedia.org/wiki/Fubini-Study metric.
\bibitem{AK}E.W. Aslaksen and J.R. Klauder, ``Continuous Representation Theory Using the Affine Group'',  J. Math. Phys. {\bf 10}, 2267-2275 (1969).
\bibitem{AK2} J.R. Klauder and E.W. Aslaksen, ``Elementary Model for Quantum Gravity", Phys. Rev. D {\bf 2}, 272-276 (1970).
\bibitem{RU2}  O.A.~Ladyzenskaja, V.~Solonnikov,  and N.N.~Ural'ceva, {\it Linear and
     Quasi-linear Equations of Parabolic Type}, (Am.~Math.~Soc., Providence, Vol.~23, 1968).
\bibitem{AK3} E W. Aslaksen and J.R. Klauder, ``Unitary Representations of the Affine Group'', J. Math. Phys. {\bf 9}, 206-211 (1968).
\bibitem{RS1} J.R. Klauder,  ``Rotationally-Symmetric Model Field Theories", J. Math. Phys. {\bf 6}, 1666-1679 (1965).
\bibitem{BCQ} J.R. Klauder, {\it Beyond Conventional Quantization} (Cambridge University Press, Cambridge, 2000).
\bibitem{UL}J.R. Klauder, ``Ultralocal Scalar Field Models", Commun. Math. Phys. {\bf 18}, 307-318 (1970).
\bibitem{UL2}J.R. Klauder, ``Ultralocal Quantum Field Theory", Acta Physica Austriaca, Suppl. {\bf VIII}, 227-276 (1971).
\bibitem{GJ}See, e.g., J.~Glimm and A.~Jaffe, {\it Quantum Physics}, (Springer Verlag, New York, 1987), Second edition.
\bibitem{DC}  D.J.E. Callaway. ``Triviality Pursuit: Can Elementary Scalar Particles Exist?'', Physics Reports {\bf 167} (5), 241–-320 (1988).
\bibitem{FW}B. Freedman, P. Smolensky, and
    D. Weingarten, ``Monte Carlo Evaluation of the Continuum Limit of $\p^4_4$ and $\p^4_3$'',  Phys. Lett. B {\bf 113}, $481-486$ (1982).
\bibitem{MA}  M. Aizenman, ``Proof of the Triviality of $\p^4_d$ Field Theory snd Some Mean-Field Features of Ising Models for $d>4$'',
Phys. Rev. Lett. {\bf 47}, 1-4, E-866 (1981).
\bibitem{JF}J. Fr\"ohlich, ``On the Triviality of $\l\s\p^4_d$ Theories and the Approach to
   the Critical Point in $d\ge4$ Dimensions'', Nuclear Physics B {\bf 200}, 281-296 (1982).
   \bibitem{PP} J.R. Klauder, ``Nontrivial Quantization of $\phi^4_n,\, n\ge2$'', Theor. Math. Phys. 182 (2015) no.1, 83-89;
Teor. Mat. Fiz. 182 (2014) no.1, 103-111; arXiv:1405.0332.
\bibitem{JS}  J. Stankowicz, private communication.
\bibitem{JK1} J.R. Klauder, ``Scalar Field Quantization Without Divergences In All Spacetime Dimensions'',
J. Phys. A: Math. Theor. 44, 273001 (30pp) (2011); arXiv:1101.1706.
\bibitem{JK2}   J.R. Klauder, ``Divergences in Scalar Quantum Field Theory: The Cause and the Cure'',
Mod. Phys. Lett. A {\bf 27}, 1250117 (9pp) (2012); arXiv:1112.0803.

\bibitem{ADM} R. Arnowitt, S. Deser, and C. W. Misner, ``The Dynamics of General Relativity'', in  {\it Gravitation: An Introduction to
Current Research}, Ed. L. Witten,
(Wiley \& Sons, New York, 1962), p. 227;  arXiv:gr-qc/0405109.
\bibitem{RK} N. Aronszajn, ``Theory of Reproducing Kernels'', Trans. Am. Math. Soc. {\bf 68}, 337 (1950).
\bibitem{TT} J.R. Klauder, {\it A Modern Approach to Functional Integration}, (Birkhauser, Boston, MA, 2010).
\bibitem{G1} J.R. Klauder, ``Noncanonical Quantization of Gravity. I. Foundations of Affine Quantum Gravity'', J. Math. Phys. {\bf 40}, 5860-5882
(1999); arXiv:gr-qc/9906013.
\bibitem{G2}J.R. Klauder, ``Noncanonical Quantization of Gravity. II. Constraints and the Physical Hilbert Space'',
J. Math. Phys. {\bf 42}, 4440-4464 (2001); arXiv:gr-qc/0102041.
\bibitem{G3} J.R. Klauder, ``Recent Results Regarding Affine Quantum Gravity'', J. Math. Phys. 53, 082501 (19pp)   (2012); arXiv:1203.0691.
\bibitem{G4}J.R. Klauder, ``Affine Quantum Gravity'', Int. Jour.
Mod. Phys. D {\bf 12}, 1769-1773 (2003); arXiv:gr-qc/0305067.
\bibitem{G5}J.R. Klauder, ``The Affine Quantum Gravity Programme", Class. Quant. Grav. {\bf 19}, 817-826 (2002); arXiv:gr-qc/0110098.
\bibitem{TH} J.R. Klauder,  ``The Action Option and the Feynman Quantization of
Spinor Fields in Terms of Ordinary $c$-Numbers", Annals of Physics {\bf 11}, 123-168 (1960).
\bibitem{BB} J.R. Klauder, ``Restricted Variations of the Quantum Mechanical Action Functional and Their Relation to Classical Dynamics",
Helv. Phys. Acta {\bf 35}, 333-334 (1962).

\end{thebibliography}
\end{document}